\documentclass[12pt,oneside,reqno]{amsart} 	% put equation numbers on the right 
\usepackage[margin=1in]{geometry}			% See geometry.pdf to learn the layout options. There are lots.
\usepackage[parfill]{parskip}				% Activate to begin paragraphs with an empty line rather than an indent
\usepackage{graphicx}
\usepackage{amssymb}
\usepackage{amsmath}					% for "split" environment
\usepackage{epstopdf}
\usepackage[hidelinks]{hyperref} %hovering over a citation number shows the citation in the bibliography, and clicking it takes you to the citation in the bibliography
\usepackage{cite}						%changes citations like [1,2,3,4,5] to [1-5] 
\title{A Time-Symmetric Resolution of the Einstein's Boxes Paradox}
\author{Michael B. Heaney\\3182 Stelling Drive\\Palo Alto, CA 94303\\mheaney@alum.mit.edu\\
Orcid ID: 0000-0002-6402-5923}
\date{10 June 2022}				% Activate to display a given date or no date
\begin{document}
\maketitle
%%%%%%%%%%%%%%%%%%%%%%%%%%%%%%%%%%%%%%%%%%%%%
\begin{abstract} 
The Einstein's Boxes paradox was developed by Einstein, de Broglie, Heisenberg, and others to demonstrate the incompleteness of the Copenhagen Formulation of quantum mechanics. I explain the paradox using the Copenhagen Formulation.~I then show how a time-symmetric formulation of quantum mechanics resolves the paradox in the way envisioned by Einstein and de Broglie. Finally, I describe an experiment that can distinguish between these two formulations.
\end{abstract}

\pagebreak
\section{Introduction} 
A grand challenge of modern physics is to resolve the conceptual paradoxes in the foundations of quantum mechanics~\cite{Smolin}.~Some of these paradoxes concern nonlocality and completeness. For~example, Einstein believed the Copenhagen Formulation (CF) of quantum mechanics was incomplete. He presented a thought experiment (later known as ``Einstein's Bubble'') explaining his reasoning at the 1927 Solvay conference~\cite{Solvay}. In~this experiment, an~incident particle's wavefunction diffracts at a pinhole in a flat screen and then spreads to all parts of a hemispherical screen capable of  detecting the wavefunction. The~wavefunction is then detected at one point on the hemispherical screen, implying the wavefunction everywhere else vanished instantaneously. Einstein believed that this instantaneous wavefunction collapse violated the special theory of relativity, and~the wavefunction must have been localized at the point of detection immediately before the detection occurred.~Since the CF does not describe the wavefunction localization before detection, it must be an incomplete theory.~In~an earlier paper, I analyzed a one-dimensional version of this thought experiment with a time-symmetric formulation (TSF) of quantum mechanics~\cite{Heaney1}, showing that the TSF did not need wavefunction collapse to explain the experimental~results.

Einstein, de Broglie, Heisenberg, and~others later modified Einstein's original thought experiment to emphasize the nonlocal action-at-a-distance effects. In~the modified experiment, the~particle's wavefunction was localized in two boxes which were separated by a space-like interval. This modified thought experiment became known as ``Einstein's Boxes.'' Norsen wrote an excellent analysis of the history and significance of the Einstein's Boxes thought experiment using the CF~\cite{Norsen}. 

Time-symmetric explanations of quantum behavior predate the discovery of the Schr\"{o}dinger equation~\cite{Tetrode} and have been developed many times over the past century~\mbox{\cite{Lewis,Eddington,Beauregard,Watanabe1,Watanabe2,Sciama,ABL,Davidon,Roberts,Rietdijk,Cramer,Hokkyo1,Sutherland,PeggBarnett,Wharton1,Hokkyo2,Miller,AV,APT,Wharton2,Gammelmark,Price,Corry, Schulman,Drummond,Heaney2,Heaney3,Heaney4}}. The~TSF used in this paper has been described in detail and compared to other TSFs before~\cite{Heaney1,Heaney2,Heaney3,Heaney4}. Note in particular that the TSF used in this paper is significantly different than the Two-State Vector Formalism (TSVF)~\cite{ABL,AV,APT}. First, the~TSVF postulates that a \textit{quantum particle} is completely described by two state vectors, written as $\langle\phi\vert\thickspace\vert\psi\rangle$, where $\vert\psi\rangle$ is a retarded state vector satisfying the retarded Schr\"{o}dinger equation $i\hbar\partial\vert\psi\rangle/\partial t=H\vert\psi\rangle$ and the initial boundary conditions, while $\langle\phi\vert$ is an advanced state vector satisfying the advanced Schr\"{o}dinger equation $-i\hbar\langle\phi\vert\partial/\partial t=\langle\phi\vert H$ and the final boundary conditions. In~contrast, the~TSF postulates that the \textit{transition} of a quantum particle is completely described by a complex transition amplitude density $\phi^\ast\psi$, defined as the algebraic product of the two wavefunctions. Second, the~TSVF postulates that wavefunctions collapse upon measurement, while the TSF has no collapse postulate.~The~particular TSF used in this paper is a type IIB model in~the classification system of Wharton and Argaman~\cite{Wharton3}.

Section~\ref{sec2} explains the paradox associated with the CF of the Einstein's Boxes thought experiment, as~described by de Broglie.~Section~\ref{sec3} reviews a CF numerical model of the thought experiment which does not resolve the paradox.~Section~\ref{sec4} describes a TSF numerical model of the thought experiment which resolves the paradox. Section~\ref{sec5} discusses the conclusions and~implications.

Note that this paper only concerns a single quantum particle interfering with itself and not multiple quantum particles entangled with each other.

%%%%%%%%%%%%%%%%%%%%%%%%%%%%%%%%%%%%%%%%%%%%%
\section{The Einstein's Boxes~Paradox}\label{sec2}
%%%%%%%%%%%%%%%%%%%%%%%%%%%%%%%%%%%%%%%%%%%%%

\noindent The Einstein's Boxes paradox was explained by de Broglie as follows~\cite{deBroglie}:

\begin{quote}
Suppose a particle is enclosed in a box $B$ with impermeable walls. The~associated wave $\psi$ is confined to the box and cannot leave it. The~usual interpretation asserts that the particle is ``potentially'' present in the whole of the box $B$, with~a probability $\vert\psi\vert^2$ at each point. Let us suppose that by some process or other, for~example, by~inserting a partition into the box, the~box $B$ is divided into two separate parts $B_1$ and $B_2$ and that $B_1$ and $B_2$ are then transported to two very distant places, for~example to Paris and Tokyo. The~particle, which has not yet appeared, thus remains potentially present in the assembly of the two boxes and its wavefunction $\psi$ consists of two parts, one of which, $\psi_1$, is located in $B_1$ and the other, $\psi_2$, in~$B_2$.  The~wavefunction is thus of the form $\psi=c_1\psi_1+c_2\psi_2$, where $\vert c_1\vert^2+\vert c_2\vert^2 = 1$.

The probability laws of [the Copenhagen Formulation] now tell us that if an experiment is carried out in box $B_1$ in Paris, which will enable the presence of the particle to be revealed in this box, the~probability of this experiment giving a positive result is $\vert c_1\vert^2$, while the probability of it giving a negative result is $\vert c_2\vert^2$. According to the usual interpretation, this would have the following significance: because the particle is present in the assembly of the two boxes prior to the observable localization, it would be immediately localized in box $B_1$ in the case of a positive result in Paris. This does not seem to me to be acceptable. The~only reasonable interpretation appears to me to be that prior to the observable localization in $B_1$, we know that the particle was in one of the two boxes $B_1$ and $B_2$, but~we do not know in which one, and~the probabilities considered in the usual wave mechanics are the consequence of this partial ignorance. If~we show that the particle is in box $B_1$, it implies simply that it was already there prior to localization. Thus, we now return to the clear classical concept of probability, which springs from our partial ignorance of the true situation. But, if~this point of view is accepted, the~description of the particle given by $\psi$, though~leading to a perfectly \textit{exact} description of probabilities, does not give us a \textit{complete} description of the physical reality, because~the particle must have been localized prior to the observation which revealed it, and~the wavefunction $\psi$ gives no information about~this. 

We might note here how the usual interpretation leads to a paradox in the case of experiments with a negative result. Suppose that the particle is charged, and~that in the box $B_2$ in Tokyo a device has been installed which enables the whole of the charged particle located in the box to be drained off and in so doing to establish an observable localization. Now, if~nothing is observed, this negative result will signify that the particle is not in box $B_2$ and it is thus in box $B_1$ in Paris. But~this can reasonably signify only one thing: the particle was already in Paris in box $B_1$ prior to the drainage experiment made in Tokyo in box $B_2$. Every other interpretation is absurd. How can we imagine that the simple fact of having observed \textit{nothing} in Tokyo has been able to promote the localization of the particle at a distance of many thousands of miles away?
\end{quote}
%%%%%%%%%%%%%%%%%%%%%%%%%%%%%%%%%%%%%%%%%%%%%

%%%%%%%%%%%%%%%%%%%%%%%%%%%%%%%%%%%%%%%%%%%%%
\section{The Conventional Formulation of Einstein's~Boxes}\label{sec3}
%%%%%%%%%%%%%%%%%%%%%%%%%%%%%%%%%%%%%%%%%%%%%

\begin{figure}
%\centering
\includegraphics[width=10 cm]{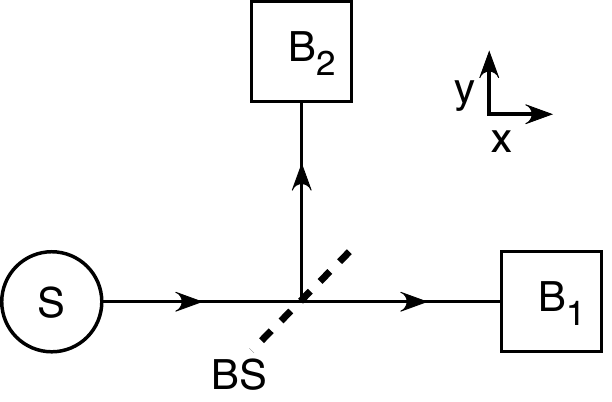}
\caption{The modified Einstein's Boxes thought experiment. The~source $S$ can emit single-particle wavefunctions on command. Each wavefunction travels to the balanced beam splitter $BS$ and then to box $B_1$ and box $B_2$. The~two boxes are separated by a space-like~interval.}
\label{fig1}
\end{figure}

The version of Einstein's Boxes proposed by de Broglie is experimentally impractical. We will use Heisenberg's more practical version~\cite{Heisenberg}, shown in Figure~\ref{fig1}. The~Conventional Formulation (CF) postulates that a single free particle wavefunction with a mass $m$ is completely described by a retarded wavefunction $\psi(\vec{r},t)$ which satisfies the initial conditions and evolves over time according to the retarded Schr\"{o}dinger equation:
\begin{equation}
i\hbar\frac{\partial\psi}{\partial t}=-\frac{\hbar^2}{2m}\nabla^2\psi.
\label{eq.1}
\end{equation}

\vspace{-6pt} 

The ``retarded wavefunction'' and ``retarded Schr\"{o}dinger equation'' are simply the usual wavefunction and Schr\"{o}dinger equation, described as retarded to distinguish them from the ``advanced wavefunction'' and ``advanced Schr\"{o}dinger equation,'' which will be defined below. We will use units where $\hbar=1$ and assume the wavefunction $\psi(\vec{r},t)$ is a traveling Gaussian with an initial standard deviation $\sigma=50$, initial momentum $k_x=0.4$, and~mass $m=1$. We will also assume that each box contains a detector whose eigenstate is the same Gaussian as that emitted by the source. The~CF assumes that a single-particle wavefunction emitted from a source $S$ will travel to the beam splitter $BS$, where half of it will pass through $BS$ and continue to box $B_1$ while the other half will be reflected from $BS$ and travel to box $B_2$. Let us assume the two halves reach the boxes at the same~time. 

\begin{figure}
%\centering
\includegraphics[width=10 cm]{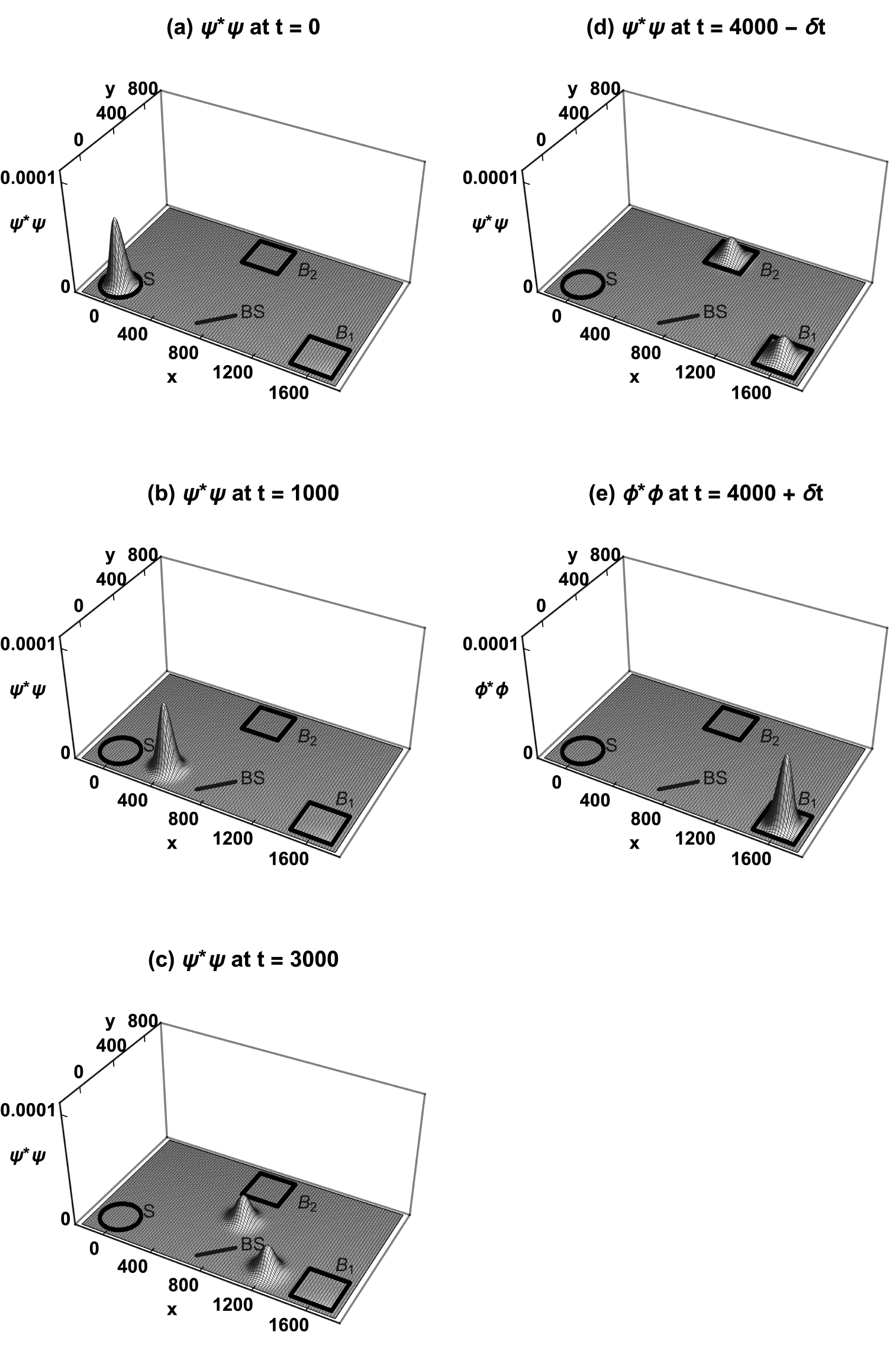}
\caption{The Conventional Formulation (CF) explanation of the Einstein's Boxes experiment with~a single-particle wavefunction emitted from source $S$.~(\textbf{a}) The probability density $\psi^\ast\psi$ is localized inside $S$. (\textbf{b}) $\psi^\ast\psi$ has left $S$ and is traveling toward the beam splitter $BS$. (\textbf{c}) $\psi^\ast\psi$ has been split in half by $BS$, and~the two halves are traveling toward boxes $B_1$ and $B_2$. (\textbf{d}) The two halves arrive at $B_1$ and $B_2$.~(\textbf{e}) A measurement at either $B_1$ or $B_2$ at $t=4000$ causes either $\psi$ to collapse to zero in $B_2$ and to a full wavefunction in $B_1$ (shown) or~$\psi$ to collapse to zero in $B_1$ and to a full wavefunction in $B_2$ (not shown).}
\label{fig2}
\end{figure}

Figure~\ref{fig2} shows how the wavefunction's CF probability density $\psi^\ast\psi$ evolves over time, assuming the initial condition is localization in the source $S$. At~$t=0$, $\psi^\ast\psi$ is localized inside the source $S$.~At~$t=1000$, $\psi^\ast\psi$ is traveling toward the beam splitter $BS$.~At~$t=3000$, $\psi^\ast\psi$ has been split in half by the beam splitter, and~the two halves are traveling toward box $B_1$ and box $B_2$.~At~$t=4000-\delta t$, half of $\psi^\ast\psi$ arrives at box $B_1$, while the other half arrives at box $B_2$.~Upon~a measurement at box $B_2$ at $t=4000$, the~CF postulates that in 50\% of the runs, the half wavefunction in box $B_2$ collapses to zero, while simultaneously, the half wavefunction in box $B_1$ collapses to a full wavefunction $\phi(\vec{r},t)$, which we will assume has the same shape and size as the initial wavefunction.~It was believed by de Broglie that this prediction of the CF was absurd: ``How can we imagine that the simple fact of having observed \textit{nothing} in [box $B_2$] has been able to promote the localization of the particle [in box $B_1$] at a distance of many thousands of miles away?''

The CF assumes that the probability $P_c$ for the collapse in box $B_1$ is $P_c=A_c^\ast A_c$, where the subscript $c$ denotes the CF and~the CF transition amplitude $A_c$ for the collapse is
\begin{equation}
A_c=\int_{-\infty}^{\infty}\phi^\ast(x,y,4000)\frac{1}{\sqrt{2}}\psi(x,y,4000)dxdy,
\label{eq.2}
\end{equation}
where $t=4000$ is the time of wavefunction collapse and~the ``quantum'' factor $\frac{1}{\sqrt{2}}$ accounts for the initial wavefunction $\psi(x,y,t)$ being split in half when it reaches box $B_1$. Plugging in numbers gives a collapse probability $P_c=0.43$. This probability is not 1/2 because the evolved wavefunction at $t=4000$ is not identical in shape to the detector eigenstate.
%%%%%%%%%%%%%%%%%%%%%%%%%%%%%%%%%%%%%%%%%%%%%
%%%%%%%%%%%%%%%%%%%%%%%%%%%%%%%%%%%%%%%%%%%%%

%%%%%%%%%%%%%%%%%%%%%%%%%%%%%%%%%%%%%%%%%%%%%
\section{The Time-Symmetric Formulation of Einstein's~Boxes}\label{sec4}

The TSF postulates that quantum mechanics is a theory about \textit{transitions} described by the transition amplitude density $\phi^\ast\psi$, where $\psi$ is a retarded wavefunction that obeys the retarded Schr\"{o}dinger equation $i\hbar\partial\psi/\partial t=H\psi$ and satisfies the initial boundary conditions, while $\phi^\ast$ is an advanced wavefunction that obeys the advanced Schr\"{o}dinger equation $-i\hbar\partial\phi^\ast/\partial t=H\phi^\ast$ and satisfies the final boundary conditions.~As~in the TSVF, $\psi$ can be interpreted as a retarded wavefunction from the past initial conditions, and~$\phi^\ast$ can be interpreted as an advanced wavefunction from the future final conditions~\cite{Heaney1}.~We will assume the same wavefunctions $\psi(\vec{r},t)$ and $\phi(\vec{r},t)$ as in the CF~above. 

An electron (e.g.) can be absorbed by a few molecules in a detector. The~number of few-molecule sites in a detector is orders of magnitude larger than the number of square centimeter sites in a detector. This makes it overwhelmingly more likely that the electron will be absorbed in an area localized to a few square nanometers than much larger areas. This could explain why the transition amplitude density refocuses to a localized area at the detector. Note that there exist two unitary solutions based on the initial conditions, but~time-symmetric theories also require the final conditions, which are that the particle is always found in either one or the other box. Let us then assume the final conditions are either a transition amplitude density localized in box $B_1$ or a transition amplitude density localized in box $B_2$.
%%%%%%%%%%%%%%%%%%%%%%%%%%%%%%%%%%%%%%%%%%%%%

\begin{figure}
%\centering
\includegraphics[width=10 cm]{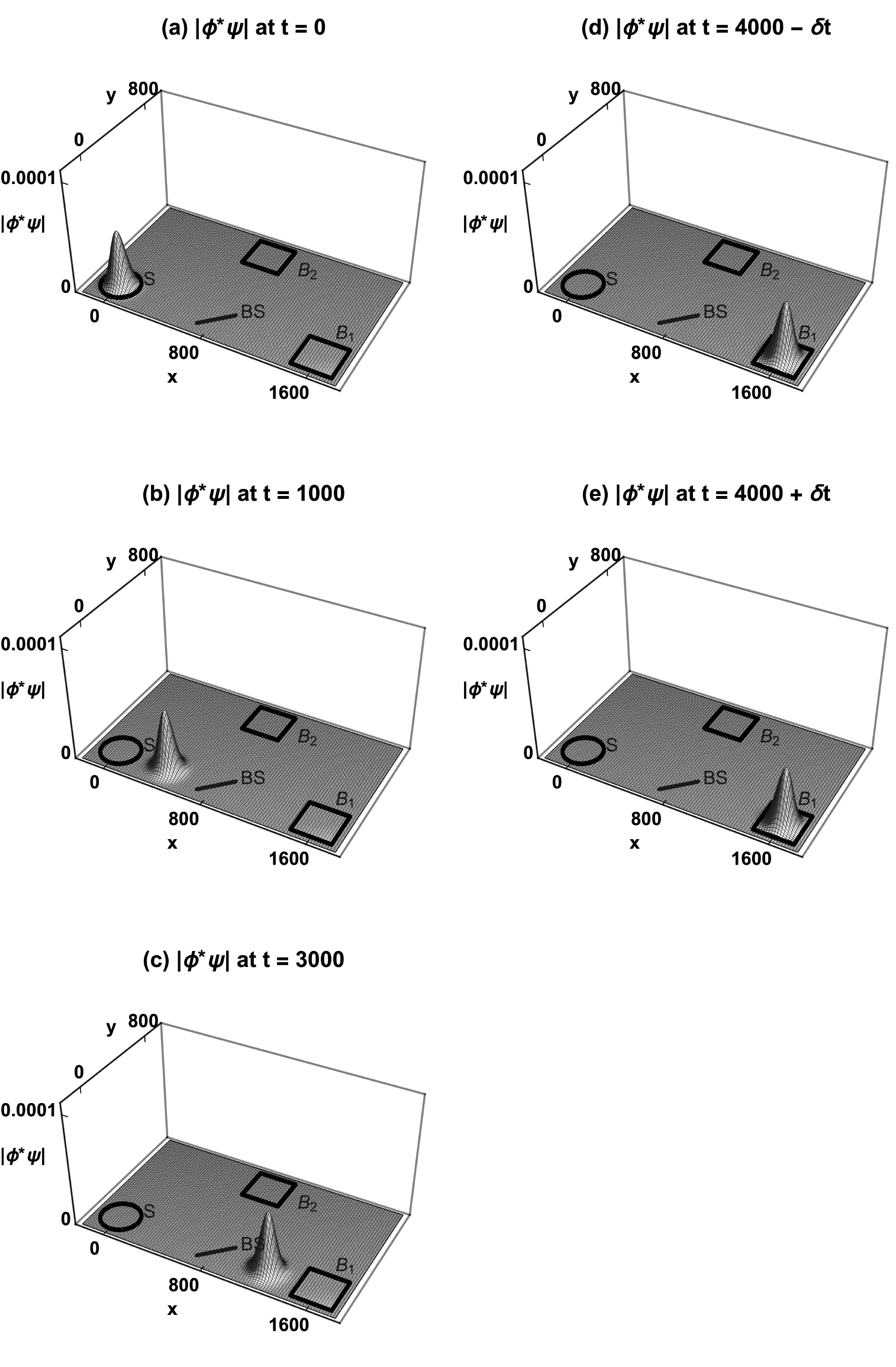}
\caption{The time-symmetric formulation (TSF) explanation of the Einstein's Boxes experiment, with~a single-particle transition amplitude density emitted from source $S$ and detected at box $B_1$. (\textbf{a}) The absolute value of the transition amplitude density $\vert\phi^\ast\psi\vert$ is localized inside $S$.~(\textbf{b}) $\vert\phi^\ast\psi\vert$ has left $S$ and is traveling toward the beam splitter $BS$.~(\textbf{c}) $\vert\phi^\ast\psi\vert$ has passed through $BS$ and is traveling toward box $B_1$. $\vert\phi^\ast\psi\vert$ is zero on the path from $BS$ to $B_2$ because $\phi^\ast$ is zero on this path. (\textbf{d}) $\vert\phi^\ast\psi\vert$ arrives at $B_1$.~(\textbf{e}) Measurements at $t=4000$ show a transition amplitude density in $B_1$ and no transition amplitude density in $B_2$.~With~equal probability, the~final condition could have been localization in box $B_2$. Transition amplitude density collapse never occurs. $\vert\phi^\ast\psi\vert$ is~normalized.}
\label{fig3}
\end{figure}

%%%%%%%%%%%%%%%%%%%%%%%%%%%%%%%%%%%%%%%%%%%%%
Figure~\ref{fig3} shows the TSF explanation of the Einstein's Boxes thought experiment, assuming that the final condition is a particle transition amplitude density localized in box $B_1$.  At~$t=0$, $\vert\phi^\ast\psi\vert$ is localized inside the source $S$. At~$t=1000$, $\vert\phi^\ast\psi\vert$ is traveling toward the beam splitter $BS$. At~$t=3000$, $\vert\phi^\ast\psi\vert$ has passed through the beam splitter and~is traveling toward box $B_1$. $\vert\phi^\ast\psi\vert$ is zero on the path from $BS$ to $B_2$ because $\phi^\ast$ is zero on this path. At~$t=4000-\delta t$, $\vert\phi^\ast\psi\vert$ arrives at box $B_1$. Upon~a measurement at box $B_2$ at $t=4000$, no particle transition amplitude density is found. Upon~a measurement at box $B_1$ at $t=4000$, one particle's transition amplitude density is found. The~one-particle transition amplitude density was localized inside box $B_1$ before the measurement was~made. 

The TSF assumes the probability $P_t$ for the transition from localization in the source $S$ to localization in box $B_1$ is $P_t=\frac{1}{2}A_t^\ast A_t$, where the subscript $t$ denotes the TSF, the~``classical'' probability factor $\frac{1}{2}$ accounts for the fact that there are two equally likely possible final states, and~the TSF amplitude $A_t$ for the transition is
\begin{equation}
A_t=\int_{-\infty}^{\infty}\phi^\ast(x,y,t)\psi(x,y,t)dxdy,
\label{eq.3}
\end{equation}

Plugging in numbers gives a TSF transition probability $P_t=0.43$, which is identical to the CF collapse probability result. Note that there is no transition amplitude density collapse in the TSF, so there is no need to specify the time of collapse in the integrand.

%%%%%%%%%%%%%%%%%%%%%%%%%%%%%%%%%%%%%%%%%%%%%
\section{Discussion}\label{sec5}
%%%%%%%%%%%%%%%%%%%%%%%%%%%%%%%%%%%%%%%%%%%%%
To the best of my knowledge, this is the first time a TSF has been shown to resolve the Einstein's Boxes paradox.~The~TSF resolves the paradox in the ways that Einstein and de Broglie envisioned. The~transition amplitude density $\phi^\ast\psi$ ``localises the particle during the propagation~\cite{Solvay},'' and $\phi^\ast\psi$ ``was already in Paris in box $B_1$ prior to the drainage experiment made in Tokyo in box $B_2$~\cite{deBroglie}.'' None of the problems associated with wavefunction collapse occur. The~TSF appears to give the sought-after exact description of the probabilities and a complete description of the physical~reality. 

One might wonder if a theory based on transition amplitude densities will be able to reproduce all of the predictions of the CF. In~1932, Dirac showed that all the experimental predictions of the CF of quantum mechanics can be formulated in terms of transition probabilities~\cite{Dirac}.~The~TSF inverts this fact by postulating that quantum mechanics is a theory which experimentally predicts \textit{only} the transition probabilities.~This implies that the TSF has the same predictive power as the~CF.

The TSF has the additional benefit of being consistent with the classical explanation of the Einstein's Boxes thought experiment. As~the size of the ``particle'' becomes larger and it starts behaving more like a classical particle, it will always go to either one box or the other. There is a logical continuity between its behavior in the quantum and classical regimes, in~contrast to the CF~predictions.

In the TSF example above, we assumed the transition probabilities for the two boxes were the same. Now consider the case where the two transitions are not equally likely. For~a very unlikely transition, the pre-experiment estimate of the TSF transition amplitude density $\phi^\ast\psi$ is tiny, while for a very likely transition, the pre-experiment estimate of $\phi^\ast\psi$ is large.~However,~this does not mean that $\phi^\ast\psi$ itself is a smaller-sized field in the event of an unlikely outcome.~Before~an experiment is conducted, we have classical ignorance of which transition will occur. We normalize the wavefunctions $\psi$ and $\phi^\ast$ to unity and calculate the expected probability for each transition based on $\phi^\ast\psi$. After~the experiment is complete, we know which of the two possible transitions actually occurred, so we renormalize the $\phi^\ast\psi$ of that transition to give a transition probability of one and~renormalize the other $\phi^\ast\psi$ to zero. Note that this is an update of our classical ignorance of which transition occurred and not a physical wavefunction collapse. This may explain why the CF collapse postulate appears to~work.

A central issue raised by the Einstein's Boxes paradox is the question of which elements of quantum theory should be thought of as elements of reality (ontic) and~which elements are merely states of knowledge (epistemic).~The~TSF transition amplitude density $\phi^\ast\psi$ and the wavefunctions $\psi$ and $\phi^\ast$ should be thought of as elements of reality, with~the understanding that $\phi^\ast\psi$ is the TSF equivalent of a real particle wavefunction while $\psi$ and $\phi^\ast$ are the TSF equivalents of virtual particle wavefunctions. For~multiple particles, $\phi^\ast\psi$ lives in a higher dimensional configuration spacetime, which should be thought of as the stage for reality \cite{Heaney3}. The~CF concept of a superposition of paths after the beam splitter then becomes just a state of knowledge in the TSF.~In reality, only one path is taken; we just do not know in advance which one. Since the TSF assumes that the sources and sinks are randomly emitting $\psi$ and $\phi^\ast$ wavefunctions, it is a probabilistic theory. In~analogy with the classical theory of special relativity, the~TSF transition amplitude density can be thought of as a quantum worldtube. The~higher dimensional configuration spacetime is then the quantum equivalent of Minkowski~spacetime.

Finally, the~CF predicts a rapid oscillating motion of a free particle's wavefunction in empty space. Schr\"odinger discovered the possibility of this rapid oscillating motion in 1930, naming it zitterbewegung~\cite{Schroedinger}.~This prediction of the CF is inconsistent with Newton's first law, since it implies a free particle's wavefunction does not move with a constant velocity. The~TSF predicts zitterbewegung will never occur~\cite{Heaney1}. Direct measurements of zitterbewegung are beyond the capability of current technology, but~future technological developments should allow measurements to confirm or deny its existence.~Given the technology, one possible way to test for zitterbewegung would be to hold an electron in the ground state in a parabolic potential and then turn off the potential while looking for radiation at the zitterbewegung frequency of $10^{21}s^{-1}$. This could distinguish between the CF and the~TSF.
%%%%%%%%%%%%%%%%%%%%%%%%%%%%%%%%%%%%%%%%%%%%%
%%%%%%%%%%%%%%%%%%%%%%%%%%%%%%%%%%%%%%%%%%%%%

\vfill
%%%%%%%%%%%%%%%%%%%%%%%%%%%%%%%
%%%%%%%%%%%%%%%%%%%%%%%%%%%%%%%%%%%%%%%%%%%%%
%%%%%%%%%%%%%%%%%%%%%%%%%%%%%%%%%%%%%%%%%%%%%
\end{document}